\documentclass[reprint,amsmath,amssymb,aip,ajp]{revtex4-2}

\usepackage{changes}
\usepackage{latexsym}
\usepackage{amsfonts}
\usepackage{mathtools}
\usepackage{bm}
\usepackage{amsthm}
\usepackage{graphicx}
\usepackage{color}
\usepackage[normalem]{ulem}
\usepackage{appendix}
\usepackage{subcaption}
\usepackage{hyperref}
\usepackage{etoolbox}

\def \vsss{\vspace{10pt}}

\def\no{\nonumber}

\def \a {\alpha}
\def \b {\beta}

\def \eps {\epsilon}

\def \Msun {M_\odot}
\def \tdot {\dot t}
\def \phidot {\dot \phi}
\def \udot {\dot u}
\def \xisco {x_{\rm ISCO}}

\def \be{\begin{equation}}
\def \bea{\begin{eqnarray}}
\def \eea{\end{eqnarray}}
\def \ee{\end{equation}}

\begin{document}

\title{The Physics presented in the Film \textit{Interstellar} and its Astrophysical Applications}

\author{Harleen Dhingra}
\email{harleen.dhingra@xaviers.edu.in}
\affiliation{St. Xavier's College, Mumbai}

\author{Sanjeev Dhurandhar}
\email{sanjeev@iucaa.in}
\affiliation{Inter University Centre for Astronomy \& Astrophysics}

\author{Sanjit Mitra}
\email{sanjit@iucaa.in}
\affiliation{Inter University Centre for Astronomy \& Astrophysics}

\date{\today}

\begin{abstract}
 
 The film {\em Interstellar} is grounded in real physics calculations. A key requirement in the film is that of a planet orbiting a supermassive black hole such that one hour on the planet corresponds to seven years on Earth. Such extreme time dilation is possible only if the planet orbits the black hole very close to its horizon. For a non-rotating (Schwarzschild) black hole, the innermost stable circular orbit (ISCO) lies at three times the Schwarzschild radius; a bound orbit between the ISCO and the event horizon is not possible. Surprisingly, general relativity allows such orbits to exist if the black hole is spinning rapidly. In this work, we present computations that are non-trivial and interesting in themselves, but more importantly, they may have useful astrophysical implications.

\end{abstract}

\maketitle

\section{Introduction}
\label{SecI}


{\em Interstellar} \cite{ThorneInterstellar}  was a science fiction film (2014) directed by Christopher Nolan. The Nobel Laureate Kip Thorne was the executive producer and science consultant for the film. Although it is a science fiction film, it is based on real calculations  \cite{ThorneInterstellar} in physics. {\em Interstellar} follows Joseph Cooper, a former NASA pilot turned farmer, who is recruited for a last-resort mission to save humanity from a dying Earth. Alongside a small crew, Dr. Amelia Brand, Dr. Romilly, Dr. Doyle, and robots TARS and CASE, he travels through a wormhole near Saturn to explore potentially habitable worlds in another galaxy. 
\par
The film envisages a planet - \textit{Miller’s Planet} - an ocean world orbiting extremely close to the supermassive black hole \textit{Gargantua} such that one hour on the planet corresponds to seven years on Earth - a huge factor of $\sim 60,000$! The crew’s brief visit on the planet results in devastating temporal consequences, as decades pass for those on Earth. The large time dilation factor arises due to the planet's orbit being very close to the horizon, which slows the passage of time relative to distant observers as explained by Einstein's Theory of General Relativity. At first sight this seems impossible, because it is well known that for a non-rotating black hole, the innermost stable circular orbit (ISCO) is three times its Schwarzschild radius, which is not at all close to the horizon and so will result in very small time dilation. But, most surprisingly, general relativistic calculations show that orbits close to the horizon can exist if the black hole is spinning sufficiently rapidly. For the required time dilation of $60,000$, the black hole needs to spin really fast - close to extremal Kerr. Gargantua itself is portrayed as a rapidly spinning (Kerr) black hole, whose rotation allows stable orbits unusually close to its event horizon. The film’s depiction of many physical effects is founded on real physics calculations, making Gargantua one of the most scientifically accurate black holes visualized in cinema.
\par

The aim of this article is to reproduce Kip Thorne's calculations in detail, which, to our knowledge, have not been presented in literature. These calculations are non-trivial and instructive in themselves, but here we argue further that they could have important implications for astrophysics.    
\par

We first show that there exist stable circular orbits skimming the horizon of a rapidly rotating black hole, which can give rise to large time dilation factors. Here we present a detailed calculation involving power-series expansions. We compare the results obtained by the power-series expansion with the numerically obtained results  from the exact equation. We further indicate how these calculations may lead to interesting astrophysics.   

\section{Time dilation}
\label{sec:time-dilation}

\subsection{Preliminaries}

We start with the basic equations describing time-like geodesics in the Kerr metric. They are most commonly written in Boyer–Lindquist coordinates $(t,r,\theta,\phi)$ which are convenient for our calculations. Setting $c = G = 1$, the metric is given by, (\citet{DhurandharMitra})
\begin{equation}
\begin{split}
ds^2 =&
-\left(1-\frac{2mr}{\Sigma}\right) dt^2
-\frac{4mar\sin^2\theta}{\Sigma} dt\, d\phi \\
&+\frac{\Sigma}{\Delta} dr^2
+\Sigma d\theta^2 \\
&+\left(r^2+a^2+\frac{2ma^2 r\sin^2\theta}{\Sigma}\right)
\sin^2\theta d\phi^2 \,,
\end{split}
\label{eq:metric}
\end{equation}
where the functions $\Sigma$ and $\Delta$ are as follows:
\begin{equation}
\Sigma = r^2 + a^2 \cos^2\theta, \hspace{2cm}
\Delta = r^2 - 2mr + a^2 \,,
\label{eq:Delta}
\end{equation}
where $m$ is the mass in length units and $a = J/m$ is the angular momentum of the black hole per unit mass of the black hole and also has dimensions of length. To obtain actual units of angular momentum, $J$ must be multiplied by the factor $G/c^3$, and $m = GM/c^2$, where $M$ is the mass of the black hole.  It is convenient to use dimensionless units for our calculations. Accordingly, we define, $x = r/m$ and $\lambda = a/m$. We keep our computations basic; our goal is to compute time-like circular orbits, in particular the ISCO, of test particles (planets) in the equatorial plane $\theta = \pi/2$. 
\par

The stationarity and axial symmetries of the Kerr metric yield two first integrals of the geodesic equations (\citet{MTW,Chandrasekhar}):
\bea
\left(1-\frac{2}{x}\right) \tdot
+ \frac{2\lambda}{x} \phidot &=& E \,, \no \\
-\frac{2\lambda}{x}\tdot
+
\left(x^2+\lambda^2+\frac{2\lambda^2}{x}\right) \phidot &=& \ell \,,
\eea
where the over dot represents derivative with respect to the dimensionless proper time $\tau/m$ of the particle. Here, $\ell = L/m$ is the dimensionless angular momentum of the particle per unit mass of the particle and black hole, at infinity; while $E$ is the energy-at-infinity of the test particle per unit mass \footnote{Since the orbits are confined to the equatorial plane, the Carter constant is zero.}. Inverting these equations for $\tdot$ and $\phidot$, we obtain,
\bea
\tdot &=&\frac{1}{\Delta}
\left[
\left(x^2+\lambda^2+\frac{2\lambda^2}{x}\right)E
-
\frac{2\lambda}{x} \ell
\right] \,, 
\label{eq:tdot}
\\
\phidot &=&
\frac{1}{\Delta}
\left[
\frac{2\lambda}{x}E
+
\left(1-\frac{2}{x}\right) \ell
\right] \,.
\label{eq:phidot}
\eea
The dimensionless version of $\Delta$ is $\Delta = x^2 - 2 x + \lambda^2$ (we denote it by the same symbol without any cause for confusion).
\vsss

The required time dilation is given by the quantity $\tdot$.

\subsection{The equations for ISCO}

From the metric Eq. (\ref{eq:metric}) and the first integrals Eqs. (\ref{eq:tdot}) and (\ref{eq:phidot}), we obtain an equation for $\dot {r}$, or equivalently, $\dot{x}$ in the equatorial plane. Following \citet{Chandrasekhar}, we set $u = 1/x$ and obtain the equation for $\udot$:
\bea
u^{-4} \dot{u}^2 &=& -\left(\lambda^2 u^2 - 2u + 1 \right) + E^2 + 2(\ell - \lambda E)^2 u^3 \, \no \\
&& - (\ell^2 - \lambda^2 E^2)u^2 \,.
\label{eq:radial}
\eea
In order to proceed towards our goal, we put $\udot = 0$. This condition only determines the turning points of the orbit.
Further, writing \cite{Chandrasekhar} $X = \ell - \lambda E$, Eq. (\ref{eq:radial}) assumes the form,
\begin{equation}
-\lambda^{2}u^{2} + 2u - 1 + E^{2}
+ 2X^{2}u^{3} - (X^{2} + 2\lambda EX)u^{2} = 0 .
\label{eq:eff_potential}
\end{equation}
In order to obtain circular orbits we need both $\dot{u} = \ddot{u} = 0$. This is equivalent to differentiating Eq. (\ref{eq:eff_potential}) with respect to $u$ and setting it equal to zero. This results in,
\begin{equation}
-2\lambda^{2}u + 2 + 6X^{2}u^{2}
- 2(X^{2}+2\lambda EX)u = 0 .
\label{eq:circ}
\end{equation}
For obtaining the ISCO, we need to further differentiate Eq. (\ref{eq:circ}) with respect to $u$. This yields,
\begin{equation}
-\lambda^{2} + 6 X^{2} u - X^{2} - 2\lambda E X = 0 \,.
\label{eq:ISCO}
\end{equation}
From Eqs. (\ref{eq:eff_potential}), (\ref{eq:circ}) and (\ref{eq:ISCO}), we obtain the following:
\bea
X^2 &=& \frac{1}{3 u^2} ~~{\rm or}~~X = \frac{1}{\sqrt{3}\,u} \,, \no \\
E^2 &=& 1 - u + X^2 u^3 \,.
\eea
We can therefore immediately solve for $E$ and $\ell$ for the ISCO (if it exists) at $u$, and so obtain, 
\begin{equation}
\begin{split}
E &= \left(1 - \frac{2}{3}u \right)^{1/2} \\
\ell &= X + \lambda E
= \frac{1}{\sqrt{3}u}
+ \lambda \left(1 - \frac{2}{3} u \right)^{1/2}.
\end{split}
\label{eq:E-ell}
\end{equation}
We may ascertain that these equations agree with the standard results for the Schwarzschild case $\lambda = 0$. Then Eq. (\ref{eq:ISCO}) gives $u = 1/6$, which is just $r = 6 m$; confirming the Schwarzschild result. Also the corresponding $\ell$ from Eq. (\ref{eq:E-ell}) is then $2 \sqrt{3}$; which is again a standard result.
\par

However, we need orbits close to the horizon in order to obtain large time dilations. This is in fact possible, as our calculations will show. Increasing $\lambda$ brings the ISCO closer to the horizon compared to the Schwarzschild case. From Eq. (\ref{eq:E-ell}), we can eliminate $E$ and $\ell$ from Eq. (\ref{eq:ISCO}). This yields a quartic in $u$ relating $u$ to $\lambda$, namely,
\be
9 \lambda^4 u^4 - 28 \lambda^2 u^3 + 6(6 - \lambda^2) u^2 - 12 u + 1 = 0 \,.
\label{eq:quartic}
\ee
We consider the extreme case of $\lambda = 1$, in which case we expect the ISCO close to the event horizon. It is easy to check that when $\lambda = 1$, $u = 1$ is a solution to Eq. (\ref{eq:quartic}). But $u = 1$ is also the horizon $r = m$ for extreme Kerr. This will imply an infinite $\tdot$ from Eq. (\ref{eq:tdot}) because $\Delta = 0$. Therefore, it makes sense to perform a power-series expansion close to the event horizon and choose $\lambda \lesssim 1$. 
Accordingly, we write $x = 1 + \eps$ and $\lambda = 1 - \delta$, where $\eps, \delta \ll 1$. We now use Eq. (\ref{eq:ISCO}) to obtain an equation in $x$, where we substitute for $E$ from Eq. (\ref{eq:E-ell}),
\begin{equation}
- \lambda^2 + 2x - \frac{1}{3}x^2
- \frac{2\lambda}{\sqrt{3}}
x^{1/2} \left(x - \frac{2}{3} \right)^{1/2} = 0
\end{equation}
In terms of $\eps$ and $\delta$, we have the equation,
\begin{equation}
\begin{split}
&- (1 - \delta)^2  + 2 (1 + \eps)
- \frac{1}{3} (1 + \eps)^2  \\
&-\frac{2}{3} (1 - \delta)
(1 + \eps)^{1/2} (1 + 3 \eps)^{1/2} = 0 \,.
\end{split}
\end{equation}
\begin{figure}[h]
    \centering
    \includegraphics[width=0.48\textwidth]{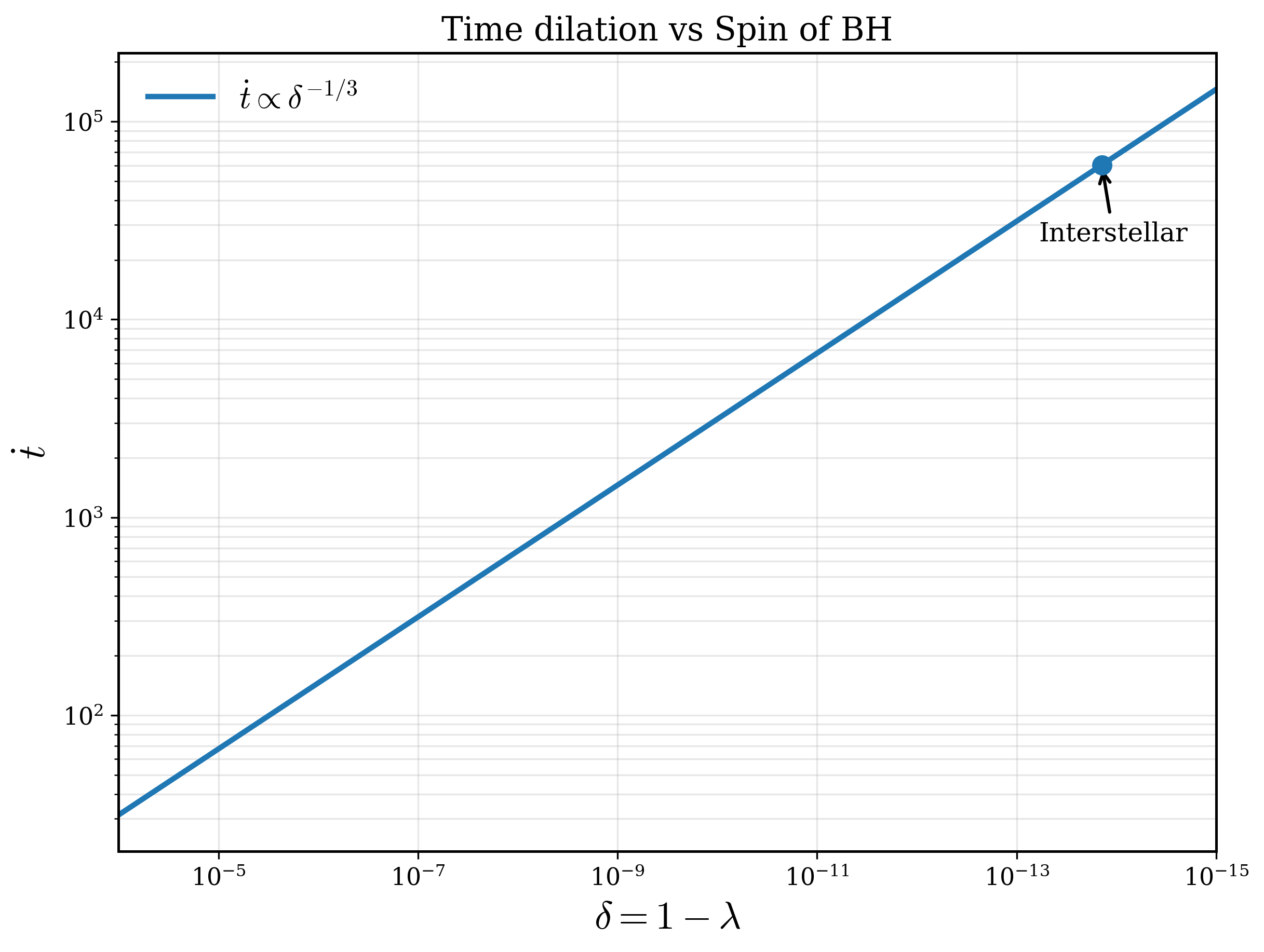}
    \hfill
    \includegraphics[width=0.48\textwidth]{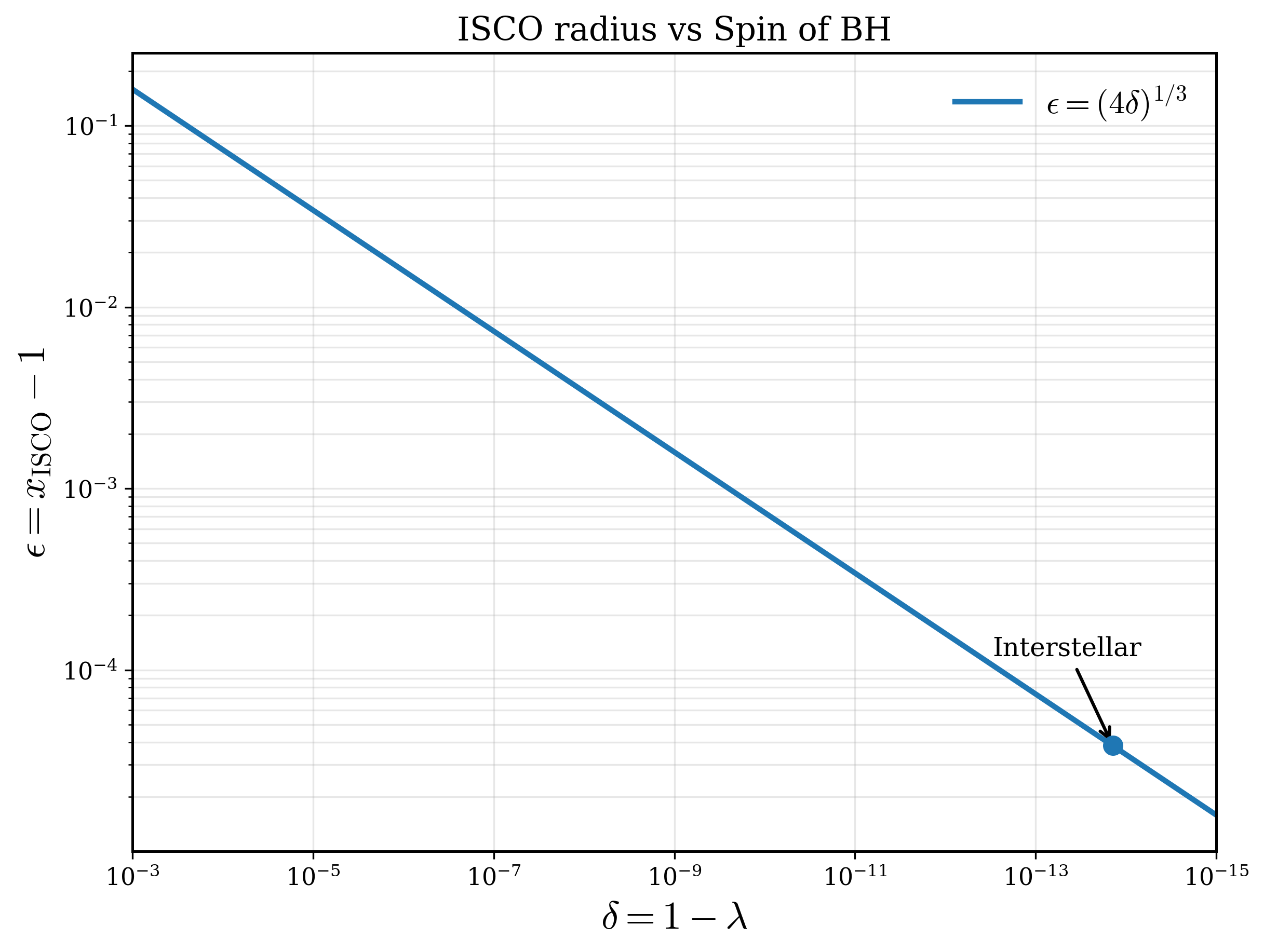}  
\captionsetup{
  justification=raggedright,
  singlelinecheck=false
}
 \caption{The figure in the upper panel shows $\tdot$ plotted versus $\delta = 1 - \lambda$. We see that as $\lambda \longrightarrow 1, ~ \tdot \longrightarrow \infty$. For Interstellar, $\tdot \simeq 60000$ and the corresponding $\delta \simeq 1.4 \times 10^{-14}$. From the figure at the bottom; the radius of the ISCO $r = (1 + \eps) m$ can be inferred from the plot of $\eps$ versus $\delta$. For {\em Interstellar}, $\eps \simeq 3.8 \times 10^{-5}$. In both plots the points for {\em Interstellar} are shown by bold dots.}
\label{fig:tdot}   
\end{figure}

We solve this equation approximately, for small values of $\eps$ and $\delta$. For this purpose, we need to perform a power-series expansion in both $\delta$ and $\eps$. One finds that for $\delta \ll \eps$; we need to expand only up to first order in $\delta$; while for $\eps$ we must keep terms up to $\eps^3$ - this is the leading term - the lower degree terms cancel out. The final result is:
\begin{equation}
\eps^3 = 4 \delta \,, ~~~~~{\rm or} ~~~~~ \eps = (4 \delta)^{1/3} \,.
\label{eq:eps_delta}
\end{equation}
The last step is to compute $\tdot$ in terms of $\delta$. To this end, we  again write out Eq.(\ref{eq:tdot}) in a convenient form,
\begin{equation}
\tdot = \frac{1}{x^2 \Delta} \left [(x^4 + \lambda^2 x^2 + 2\lambda^2 x) E - 2\lambda \ell x \right] \,,
\label{eq:tdot2}
\end{equation}
and write $E$ and $\ell$ in terms of $\eps$. We need these quantities only up to the first order in $\eps$. We readily obtain from Eq. (\ref{eq:E-ell}), $E = (1 + \eps)/\sqrt{3}$ and $\ell = 2 (1 + \eps)/\sqrt{3}$. Also to the required order, $\Delta \approx \eps^2$. Substituting these expressions into Eq. (\ref{eq:tdot2}), we obtain,
\begin{equation}
\dot t = \frac{4}{\sqrt{3}} \frac{1}{\epsilon} = \frac{4^{2/3}}{\sqrt{3}} \delta^{-1/3} ~~~~{\rm or} ~~~~~ \delta =
\frac{16}{3\sqrt{3}}
\frac{1}{(\dot t)^3} \,.
\label{eq:delta}
\end{equation}

In {\em Interstellar}, one hour on the Miller's planet corresponds to approximately seven years on Earth, due to time dilation. This is about a factor of $60,000$. Using this value in Eq. (\ref{eq:delta}), we obtain $\delta \simeq 1.4 \times 10^{-14}$. Thus, the black hole Gargantua is spinning very fast; extremely close to extreme Kerr. We display in Fig.~\ref{fig:tdot} two plots. The plot on the top  shows $\tdot$ versus $\delta$ while the position of the ISCO $r = (1 + \eps)m$ can be inferred from the plot of $\eps$ versus $\delta$ at the bottom. The points corresponding to {\em Interstellar} are marked by bold dots in both figures.

\section{Beyond {\em Interstellar}}

 The computations in section \ref{sec:time-dilation} have implications to interesting astrophysics. 
 \par
 
 One question that may arise is whether Miller’s planet would be torn apart by the tidal forces of the black hole. The mass of the black hole Gargantua is taken to be very large $\sim 10^8 \Msun$. This implies that just outside the horizon, the gravitational field is quite weak and the tidal forces are not very strong. It is argued in \citet{Luminet2015} that the planet remains marginally intact - the Roche limit is of the same order as the size of the horizon of the black hole $\sim 10^8$ km.

 \subsection{Accuracy of the power-series approximation}
 
\begin{figure}[h]
\centering
    \includegraphics[height=5cm, keepaspectratio]{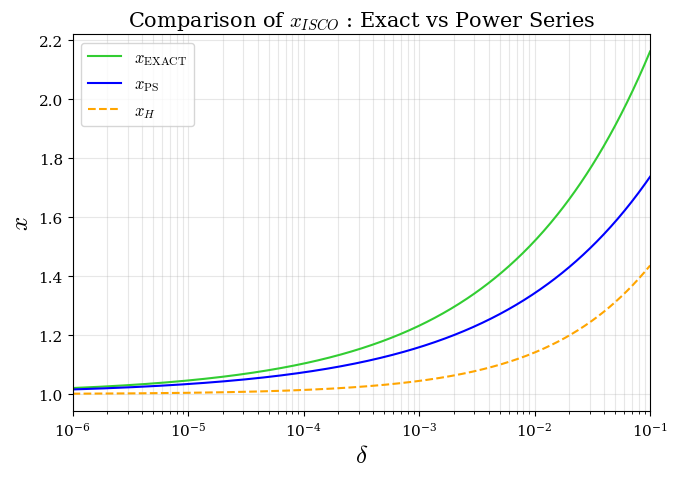}
    \hfill
    \includegraphics[height=5cm, keepaspectratio]{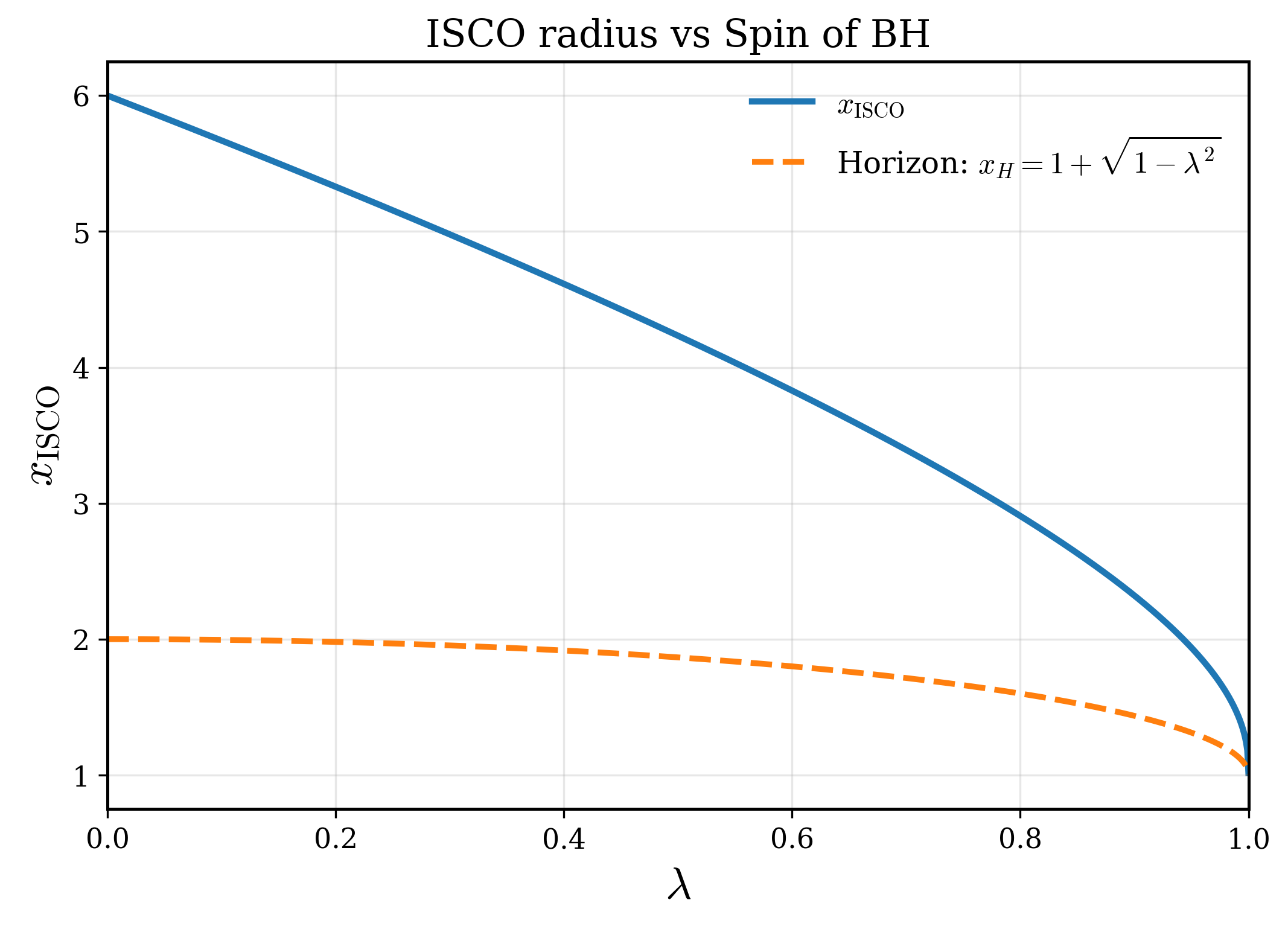}
\captionsetup{
  justification=raggedright,
  singlelinecheck=false
}
 \caption{The upper panel shows the exact value of the radius of ISCO obtained numerically and that obtained by the power-series approximation. The power-series approximation gives the ISCO closer to the horizon than the exact case. The two methods give essentially the same result when $\delta \lesssim 10^{-4}$. The horizon is also shown for reference. The bottom panel displays the radius of the ISCO as a function of $\lambda$ as it ranges from $0$ to $1$. The ISCO then reduces from $r = 6m$ for Schwarzschild to $r = m$ for extreme Kerr.}
\label{fig:x_isco}   
\end{figure}

It is incumbent on us to investigate how accurately the approximation we have made in Eq. (\ref{eq:eps_delta}) is valid. In order to achieve this, we may solve Eq. (\ref{eq:quartic}) numerically. The results are shown in the upper panel of Fig. \ref{fig:x_isco}. From the figure it is evident that the exact ISCO is further from the horizon than what the power-series approximation implies. When $\delta \lesssim 10^{-4}$, the difference in the ISCO radii computed  by the two methods is less than 1\%. 

\subsection{Redshift of photons emitted by accretion discs}

A compelling astrophysical  application is that of accretion discs around black holes. In {\em Interstellar}, the black hole is extremely close to extreme Kerr ($\lambda = 1$) with $\delta \simeq 10^{-14}$. But in the usual astrophysical scenarios, the $\delta \gtrsim 10^{-3}$. To this end, in the bottom panel of Fig. \ref{fig:x_isco}, we plot $\xisco$ versus $\lambda$ for the full range $0 \leq \lambda \leq 1$, which is from Schwarzschild to extreme Kerr. For reference, we also plot the horizon $x_{\rm h} = 1 + \sqrt{1 - \lambda^2}$ to show how the ISCO orbit approaches the horizon as $\lambda$ increases from zero to unity. This has important implications for the red-shift, because the red-shift increases as the orbiting material of the accretion disc approaches the horizon. The spectral lines, for example; the Fe(Iron) $K_{\alpha}$ lines , will display broadening because the orbits are at different distances from the horizon. For a fast spinning black hole, this broadening would be quite large as our results indicate. The broadening of the spectral lines can be used to estimate the spin of the black hole.
\par

\begin{figure}[h]
\centering
    \includegraphics[height=5cm, keepaspectratio]{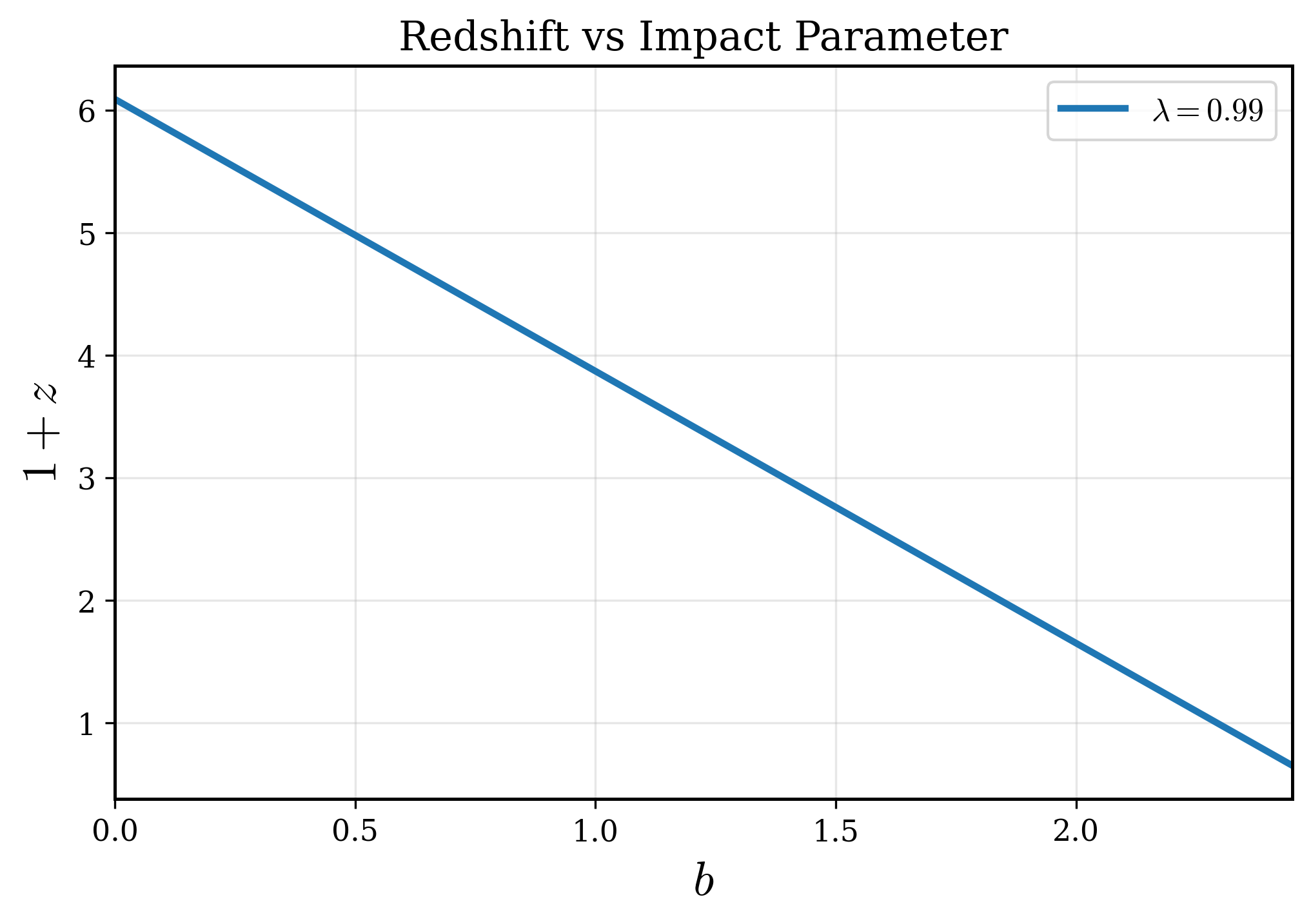}
    \hfill
    \includegraphics[height=5cm, keepaspectratio]{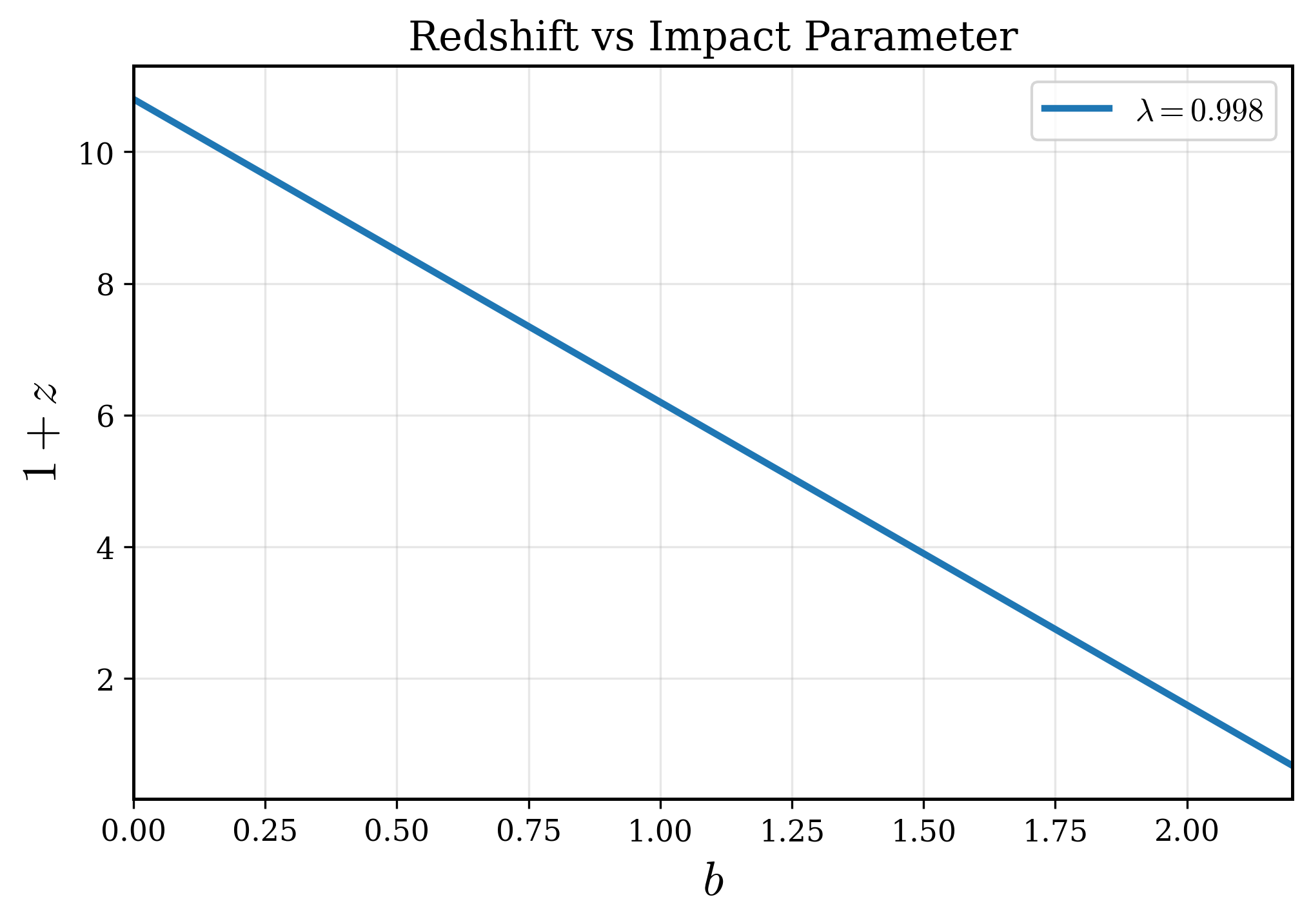}

\captionsetup{
  justification=raggedright,
  singlelinecheck=false
}
\caption{The figures show redshift as a function of $b$, where $0 \leq b \leq b_{\rm max}$ for $\lambda = 0.99$ (top) and $0.998$ (bottom).}    
\label{fig:redshift}
\end{figure}

The equations for null geodesics are essentially the same as Eqs. (\ref{eq:tdot}) and (\ref{eq:phidot}), where now the proper time is replaced by an affine parameter $\sigma$, and there is only one parameter, $b = \ell/E$, the impact parameter, which characterizes the null geodesics, instead of the two parameters $E$ and $\ell$ which characterize time-like geodesics. The relevant first integrals are: 
\bea
\frac{dt}{d \sigma} &=&\frac{1}{\Delta}
\left[
\left(x^2+\lambda^2+\frac{2\lambda^2}{x}\right)
- \frac{2\lambda}{x} b \right] \equiv k^t \,, \no \\
\frac{d \phi}{d \sigma} &=&
\frac{1}{\Delta}
\left[\frac{2\lambda}{x} + \left(1-\frac{2}{x}\right) b \right] \equiv k^\phi \,.
\label{eq:phidot_null}
\eea
We denote the wave vector by $k^\a$, where $k_\a k^\a = 0$ is a null vector. We consider the geodesics in the equatorial plane, so $k^\theta = 0$ and since we will be considering near extremal Kerr black holes, we put $\lambda = 1$ in the equation. Then the radial component satisfies:
\be
\left (\frac{dx}{d \sigma} \right)^2 = 1 + \frac{2}{x^3} (b - 1)^2 - \frac{1}{x^2} (b^2 - 1) \,.
\label{eq:rad_null}
\ee
For motion to be possible the RHS of the above equation should be non-negative. Since we would like the photon emitted from the accretion disc to be observed by an asymptotic observer; we must choose $dx/d\sigma \geq 0$ in Eq.(\ref{eq:rad_null}). Or writing the RHS of this equation in powers of $b$, it  must obey,
\be
(2 - x) b^2 - 4b + x^3 +x + 2 \geq 0 \,.
\label{eq:b}
\ee
The relevant root of this equation, when the LHS of Eq.~(\ref{eq:b}) is set equal to zero is $b = 1 + x$. Since we will choose $x$ to be the radius of the ISCO,  $x \gtrsim 1$. In order that Eq.~(\ref{eq:b}) be satisfied, we must choose $b \leq b_{\rm max} = 1 + \xisco$. The redshift is computed by the formula:
\be
1 + z = \frac{(g_{\a \b} u^\a k^\b)_{\rm e}}{(g_{\a \b} u^\a k^\b)_{\rm o}} \,,
\label{eq:z}
\ee
where the subscripts ``e" and ``o" denote the quantities at emission and observation. Since, we will assume the observer to be asymptotic and static, we have for the observer, only $u^t = 1$, while the space components $u^i = 0$. Also at infinity, $k^t \longrightarrow 1$ and $g_{tt} \longrightarrow 1$. Thus, the denominator in Eq. (\ref{eq:z}) is just unity; only the numerator needs evaluation where the quantities must be evaluated at the ISCO. Since $u^\a$ has only $u^t$ and $u^\phi$ non-zero, only these components of $k^\b$ are required.  Thus,
\begin{equation}
\begin{split}
1 + z =&\,
g_{tt} u^t k^t
+ g_{t \phi} (u^t k^\phi + u^\phi k^t) \\
&+ g_{\phi \phi} u^\phi k^\phi
\equiv u^t - b u^\phi \,.
\end{split}
\label{eq:redshift}
\end{equation}
As seen in the above equation, for the ISCO, the expression for the redshift simplifies considerably. From astrophysical considerations, $\lambda \lesssim 0.998~$ (\citet{Luminet2015}). Given $\lambda$, we can compute both $u^\a$ and $k^\a$ at the ISCO from the previous equations. Suppose we choose $\lambda \simeq 0.99$, the $\xisco \sim 1.45$ then, we must choose $b \leq 2.45$. Suppose, we choose $b = 1$, then $1 + z \sim 3.86$, while if we choose $b = 0$, the redshift factor $\sim 6.09$. The reason for this is the Doppler effect; for $b = 1$, the photon is emitted essentially in the direction of the velocity of the particle, while when $b = 0$, it is emitted radially and the gravitational redshift dominates. Below, in Fig. \ref{fig:redshift} we have plotted the redshift versus $b$ from $0 \leq b \leq b_{\rm max}$ for two values of the black hole spin $\lambda = 0.99$ and $\lambda = 0.998$. From the figures we see that when $b$ is close to $b_{\rm max}$, there is a net blue-shift - the Doppler effect dominates the gravitational.

 Let us now turn to the period of the ISCO and the time dilation factor. The time dilation factor is just $u^t$, which from Eq. (\ref{eq:redshift}), is the same as the redshift for the impact parameter $b = 0$. Thus, for the values chosen  $\lambda = 0.99, 0.998$, from the Fig. \ref{fig:redshift}, the time dilation factors can be read off and are $\sim 6.1$ and $\sim 11$ respectively. For computing the ISCO period, we require the mass of the black hole. We choose this to be the same as that of Gargantua, $M_{\rm BH} \sim 10^8 \Msun$. The period of ISCO is given by $(2 \pi/u^\phi) \times G M_{\rm BH}/c^3 $. For the  values of $\lambda$ chosen, the periods are $\sim 1400$ sec. and $\sim 670$ sec. respectively. These calculations have implications for gravitational waves (GW) emitted by a stellar mass black hole binary orbiting a supermassive black hole.
 
\section{Concluding Remarks}
\label{SecIV}

In this work we have provided explicit calculations pertaining to the time dilation factor, which is very large $\sim 60,000$ as required in {\em Interstellar}. Such large time dilation factors are only possible if orbits of particles (planets) can exist close to the event horizon of the black hole. It is shown that such orbits do exist if the black hole is spinning close to extremal Kerr. In particular, for a time dilation factor of $\sim 60000$, the dimensionless black hole spin $\lambda$ is given by $1 - \lambda \sim 10^{-14}$, and the ISCO skims the event horizon. We also compared the power series approximation with numerically obtained results for the radius of ISCO. Further, we have performed the computations for the redshift of photons emitted by a particle on ISCO.  More importantly, we argue that these calculations have relevance to astrophysics, such as accretion disks and GW emitted by compact binaries orbiting close to supermassive black holes. We elaborate on these points below.  
\par

While this work reproduces the results obtained in the context of the film {\em Interstellar}, its application can be useful in broader areas of astrophysics, especially those related to phenomena near black holes. For instance, radiation emitted from the accretion disk around black holes can get significantly redshifted near the event horizon. The profile of the observed emission lines, namely the Iron $K_{\alpha}$ lines, therefore become a convolution of the intensity distribution and redshift (gravitational and Doppler) across the accretion disks. Rigorous, yet highly involved, calculations of such line profiles, albeit numerical, have been provided in literature \cite{MartocchiaKarasMatt,BromleyMillerPariev,Laor}.  However, the calculations presented in our work can provide a more straightforward way to obtain physical insights, even if they are not highly precise. 
\par

Another such application could be in the context of GW emission from compact binaries (\citet{Creighton}). It is quite possible that around supermassive black holes at the galactic centers, there is a higher concentration of much lighter inspiraling or merging binaries (\citet{Yang}). The GW signal from these binaries can also get highly redshifted when they are close to the event horizon, which can shift their frequencies to a much lower frequency band; for example, from tens of Hz for a stellar mass binary, that is observable in the LIGO \cite{LIGO} band to the deciHz band; or from a deciHz band for an intermediate-mass blackhole binary to the mHz band observable by LISA \cite{LISA2017}. Also, the various modes such as w, p and f modes \cite{KostasBernard92,Andersson_Kokkotas1998} emitted by neutron stars in the kHz band would be redshifted into the sensitive band in the range of few hundred Hz of the current ground-based GW detectors. The time scale on which a typical black hole binary merges in ground based detectors, is of the order of few seconds, while our computations here show that for $\lambda \sim 0.99, 0.998$, the periods are several hundred seconds so that the binary would merge almost at the same phase of the orbit, which would relatively simplify the calculation of the GW waveform. If on the other hand, the binary merger timescale is of the same order as the ISCO period, then the GW waveform would be modified in a complex manner by time-dependent Doppler effects and time-changing polarisation. While obtaining the exact waveform in such situations can be highly involved; based on the geometry and evolution of the trajectory of the center of mass of the binary around the supermassive black hole, our calculations therefore provide a relatively straightforward way to predict what may be expected in such situations and whether a full-fledged investigation would be worth pursuing.

\section{Acknowledgments}

This work is based in part on the M.Sc. dissertation of Harleen Dhingra submitted to St. Xavier's College, Mumbai, India. SVD would like to thank Sajal Mukherjee for clearing up an issue. The authors would like to thank Shubhajeet Das and Sayan Kar for valuable comments that helped in the re-examination of the ISCO analysis. 

\section{Bibliography}

\bibliographystyle{aipnum4-2}
\bibliography{refs}

\end{document}